# Integrated broadband bowtie antenna on transparent substrate

Xingyu Zhang*,[a], Shiyi Wang[b], Harish Subbaraman[c], Qiwen Zhan[b], Zeyu Pan[a], Chi-jui Chung[a], Hai Yan[a], and Ray T. Chen*,[a,c]

[a] University of Texas at Austin, 10100 Burnet Rd, MER 160, Austin, TX 78758, USA;
[b] University of Dayton, 300 College Park, Dayton, OH 45469-2951, USA;
[c] Omega Optics, Inc., 8500 Shoal Creek Blvd, Austin, TX 78757, USA.

## ABSTRACT

The bowtie antenna is a topic of growing interest in recent years. In this paper, we design, fabricate, and characterize a modified gold bowtie antenna integrated on a transparent glass substrate. We numerically investigate the antenna characteristics, specifically its resonant frequency and enhancement factor. We simulate the dependence of resonance frequency on bowtie geometry, and verify the simulation results through experimental investigation, by fabricating different sets of bowtie antennas on glass substrates utilizing CMOS compatible processes and measuring their resonance frequencies. Our designed bowtie antenna provides a strong broadband electric field enhancement in its feed gap. The far-field radiation pattern of the bowtie antenna is measured, and it shows dipole-like characteristics with large beam width. Such a broadband antenna will be useful for a myriad of applications, ranging from wireless communications to electromagnetic wave detection.

## 1. INTRODUCTION

In recent years, bowtie antennas have been a hot topic for intense theoretical and experimental investigation, impelled by its unique features and advantages, including simple planar structure, stable broadband performance, strong near-field enhancement [1-13]. One simple configuration used to achieve broadband characteristics is a biconical antenna formed by two cones, and the bowtie antenna is a simplified two-dimensional structure using two triangular shapes separated with a small gap that resembles the shape of a bowtie [14] [15]. The bowtie antenna concentrates the energy and provides a good localization of electric field inside the feed gap, providing a strong near-field enhancement [16, 17], which is useful for several applications, including optical sensing and energy harvesting [2, 18-20]. In addition, the resonant frequency of a bowtie antenna can be designed and tuned by appropriately modifying and scaling the bowtie geometry, such as the arm length, the flare angle, and the feed gap width. This enables the bowtie antenna to have various applications over a wide frequency range, including extreme-ultraviolet light generation [1], local optical absorption [2], Terahertz-wave near-field imaging [3], mid-infrared plasmonic antennas [4], microwave radar [5], wireless communications [6], and flexible RF devices [7].

A modified gold bowtie antenna on lithium niobate (LiNbO3) substrate has been theoretically studied previously for microwave photonic applications [21]. The structure of this antenna is a conventional bowtie shape with extension bars attached to its apex points. Under RF illumination, an extended near-field area with a uniformly enhanced local electric field is obtained in its feed gap. Recently, such an optimized bowtie antenna has been integrated with an electro-optic modulator in its feed gap to form a very sensitive electromagnetic wave sensor [22, 23]. For the detection of electromagnetic waves, the interaction of the waves and the substrate materials needs to be considered, and a low-k dielectric substrate, such as a glass substrate, is desired to provide better microwave coupling [23]. In addition, some electromagnetic wave detectors previously demonstrated on silicon-on-insulator (SOI) substrates [22, 24] suffer from unwanted reflection and scattering from backside silicon handle [25], while, in comparison, a glass substrate can avoid this issue and improve the detection sensitivity. Therefore, a cost-effective transparent low-k glass substrate is used in this work.

In order to provide a guide for further development and better optimization of this type of bowtie antenna on glass substrate, more detailed theoretical and experimental investigations on this type of bowtie antennas with different geometrical parameters are required. Especially, experiments need to be performed to verify the theoretical analysis. In this paper, we design, fabricate and experimentally characterize a modified gold bowtie antenna on transparent glass substrate targeted around an operating frequency of 10.5GHz. The designed bowtie antenna is miniaturized with an area smaller than 1cm$^2$. The electric-field enhancing capability of the bowtie antenna is investigated by numerical simulations. Its geometry-dependent resonant frequency is simulated, and then it is experimentally verified on a group of fabricated bowtie antennas on glass substrate. The radiation pattern of this bowtie antenna is also measured.

*xzhang@utexas.edu; phone 1 512-471-4349; fax 1 512 471-8575
*raychen@uts.cc.utexas.edu; phone 1 512-471-7035; fax 1 512 471-8575

## 2. DESIGN

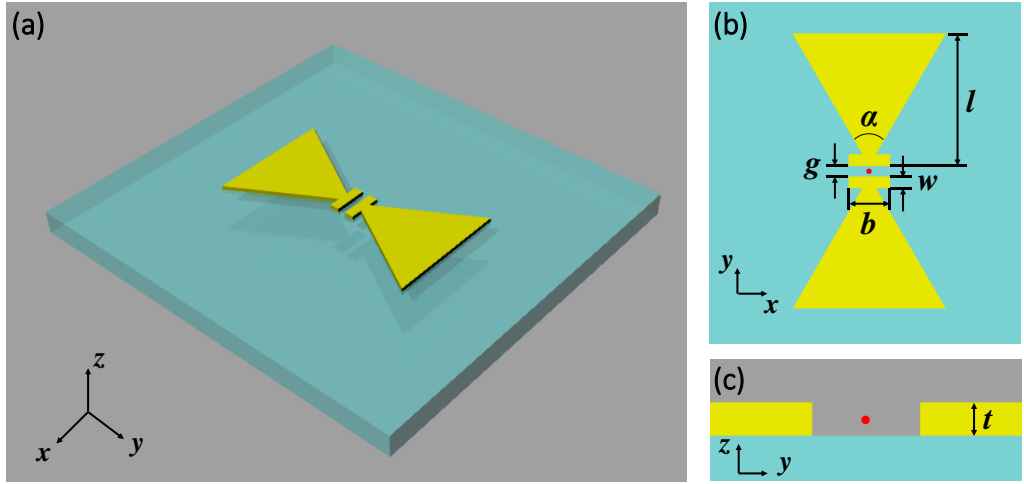

Figure 1. (a) 3D perspective of our modified gold bowtie antenna on a transparent glass substrate. (b) Top view of the bowtie antenna. $l$: arm length; $\alpha$: flare angle; $g$: feed gap; $w$: bar width; $b$: bar length. (c) Cross-sectional view of the bowtie antenna. $t$: antenna thickness. The red dot in the center of the feed gap at middle height indicates the observation point for the simulation of electric field enhancement.

A schematic of the modified bowtie antenna on glass substrate is shown in Fig. 1, consisting of a conventional bowtie antenna with capacitive extension bars attached to the apex points of the bowtie [21]. The extension bars have a length, $b$=300μm, a width, $w$=10μm and a feed gap, $g$=10μm. The thickness of the gold film is chosen to be $t$=5μm, which is far beyond the skin depth of gold at the RF frequency of operation. The thickness of glass substrate is 1mm. Under RF illumination, highly enhanced local electric field is generated inside the feed gap of this bowtie antenna. The current on the bowtie antenna surface, induced by incident RF field, charges the feed gap and subsequently establishes this strong electric field in the feed gap [26]. Generally, the antenna system can be considered as a typical LC circuit, which is mainly composed of the inductive bowtie metallic arms and the capacitive bars, giving rise to an LC resonance determined by the antenna geometry. In this work, this resonance effect is characterized by field enhancement factor, defined as the resonant electric field amplitude at a specific observation point [red dot in Figs. 1 (b) and (c)] divided by the incident electric field amplitude. With the feed gap ($g$=10μm) and capacitive bars ($b$=300μm, and $w$=10μm) fixed, the resonant frequency of a bowtie antenna is mainly determined by the length of each bow arm and the flare angle [$l$ and $\alpha$ in Fig. 1 (b)] [27]. COMSOL Multiphysics is used to simulate this bowtie antenna model. The incident electric field is a normalized continuous plane wave linearly polarized along the antenna axis (y-direction) and impinges upon the antenna from the top. Simulation results show that, with the bow arm length $l$=5.5mm and the flare angle $\alpha$ =60°, the bowtie antenna has a maximum field enhancement around a resonant frequency of 10.5GHz, and a uniform broadband electric field enhancement over the entire feed gap is created. Fig. 2 (a) shows the simulated field enhancement spectrum of this bowtie ($l$=5.5mm, $\alpha$ =60°) as a function of frequency, indicating that the electric field radiation compressed inside the feed gap is enhanced by a maximum factor of ~670 at 10.5GHz, with a 1-dB RF bandwidth over 9GHz. Figs. 2 (b) and (c) show both the top view and the side view of the simulated normalized local electric field amplitude at the resonant frequency. The polarization of the electric field is along the x direction in Figs. 2 (b) and (c). The electric field is mainly confined in the feed gap region, and the peak resonant frequency can be tuned by adjusting the arm length and the flare angle of the bowtie antenna. The simulated resonant frequencies at different arm lengths and flare angles are shown in Figs. 6 and 7, respectively, and will be correlated with experimental results in Section IV. More detailed theoretical analysis of the geometry-dependent performance of bowtie antennas can be found in Refs. [9, 17, 28].

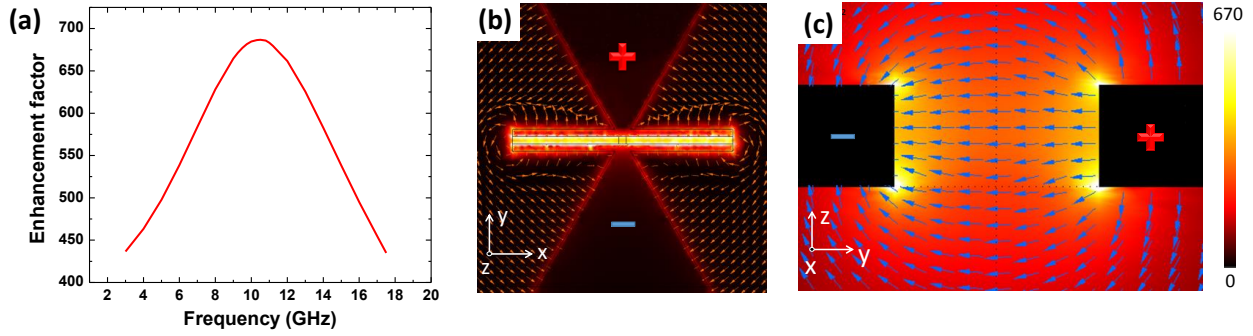

Figure 2. (a) The simulated field enhancement spectrum for a bowtie antenna with the arm length of 5.5mm and the flare angle of 60º, indicating a field enhancement factor of ~670 at 10.5GHz and a 1-dB RF bandwidth over 9GHz. (b) Top view of the simulated normalized electric field enhancement distribution at the resonant frequency. (c) Cross-sectional view of the simulated electric field enhancement distribution at the resonant frequency.

## 3. FABRICATION

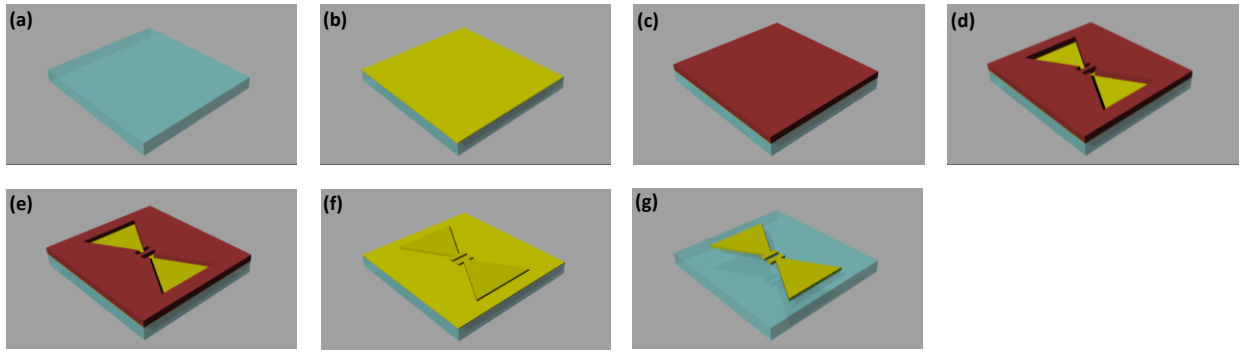

Figure 3. Fabrication process. (a) A glass substrate. (b) Seed layer deposition. (c) Photo resist spincoating. (d) Photolithogprahy. (e) Electroplating. (f) Photo resist removal. (g) Seed layer removal.

Two groups of bowtie antennas are fabricated on 1mm-thick glass substrates (Fisher Scientific 12-550C) through standard CMOS manufacturing process. The first group of five bowtie antennas is fabricated with a fixed flare angle of 60 degrees, but arm lengths varying from 3.5mm to 5.5mm in steps of 0.5mm. The second group of three bowtie antennas has a fixed arm length of 4.5mm, but flare angles of 30 degrees, 60 degrees and 90 degrees, respectively. A schematic process flow of fabrication is shown in Fig. 3. A 50nm-thick gold seed layer with a 5nm-thick chromium adhesion buffer is deposited on the glass substrate by electron-beam evaporation. A 10 μm-thick AZ-9260 photoresist is spincoated on the seed layer and baked at 90ºC for 2 min and then at 110ºC for 2min. A buffer mask for the bowtie structure is patterned by photolithography (dose: 260mJ/cm$^2$) and AZ400K developer (diluted 1:4). Next, a 5 μm-thick gold film is electroplated by through-mask plating method in a magnetically stirred neutral noncyanide electrolyte (Techni-Gold 25ES) under a constant current of 8mA at the temperature around 50ºC. After electroplating, the buffer mask is removed using Acetone and then the gold seed layer is removed using wet etchant, leaving the electroplated gold bowtie structure on top of the transparent glass substrate. The conductivity of the electroplated gold film is measured to be $2.2 \times 10^7$S/m. Microscope images of a few fabricated devices are shown in Fig. 4.

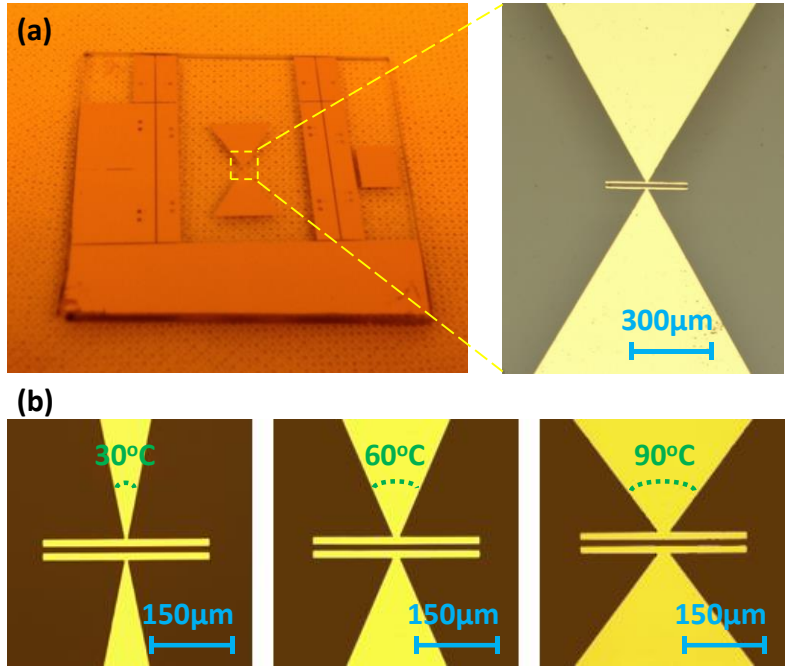

Figure 4. (a) A fabricated bowtie antenna on glass substrate. (b) Three bowtie antenna with flare angles of 30 degrees, 60 degrees and 90 degrees, respectively. For all these three bowtie antennas, the arm length *l*=4.5mm, the extended bar length b=300μm, extended bar width w=10μm, and the feed gap g=10μm.

## 4. CHARACTERIZATION

First, a fabricated bowtie antenna with arm length of 5.5mm and flare angle of 60° is tested as a receiving antenna. RF signal at 10.5GHz from a vector network analyzer (HP 8510C) is applied to an X-band horn antenna which is mounted on top of the fabricated bowtie antenna. The horn antenna is place sufficiently away in its far-field for the assumption of quasi-plane wave to hold. The electromagnetic power that the bowtie antenna receives is measured by a microwave spectrum analyzer (HP 8560E) via a ground-signal (GS) microprobe (Cascade Microtech ACP40GS500) which contacts the bow arms of the bowtie. Fig. 5 shows the received power of the bowtie antenna. The power response is about 30dB above the noise floor at 10.5GHz. This simple test demonstrates the functionality of the fabricated bowtie antenna. Due to the reciprocity, the bowtie antenna should also work well as a transmitting antenna, which will be demonstrated in the next test.

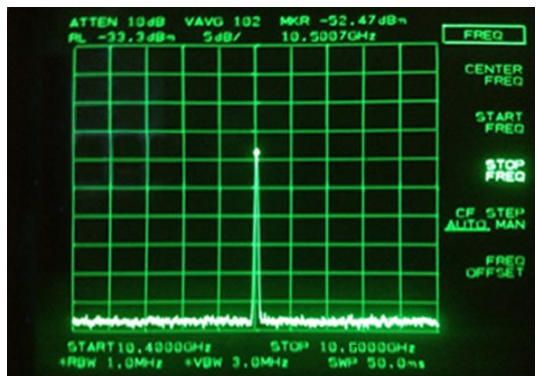

Fig. 5. Measured response power of the bowtie antenna as a receiving antenna at 10.5GHz.

Next, in order to demonstrate the broadband characteristics of the fabricated bowtie antenna and to investigate the dependence of resonant frequency on the bowtie geometry, the two groups of bowtie antennas are tested as transmitting antennas. The vector network analyzer is used to measure the $S_{11}$ parameter (reflection signal) of these antennas over a broad frequency range of 1-20GHz via the GS microprobe. Assuming negligible loss, the normalized transmission signal can be inferred from the $S_{11}$ measurements, as shown in Figs. 6 (a) and (c), from which a broadband response can be clearly seen. The measured transmission signals of the first group of bowtie antennas are shown in Fig. 6 (a). The measured resonant frequency as a function of arm lengths is extracted from the figure, and then correlated with the simulated resonant frequency, as shown in Fig. 6 (b). Similarly, the measured resonant frequency as a function of flare angles is extracted from the measured transmission signals of the second group of bowtie antennas in Fig. 6 (c), and then correlated with simulated resonant frequency in Fig. 6 (d). For longer bowtie arm or larger flare angle, the current flows through longer path to the gap, so the effective antenna size is increased, leading to longer resonant RF wavelength, which corresponds to lower resonant frequency. The trend in measurement results agree with the simulations. It can be seen that there are still some deviations between the measured and simulated resonant frequencies. This could be due to several reasons, such as the difference of the dielectric constant of an actual glass substrate and that assumed in simulations, and slight variations of size and shape of fabricated bowtie antenna from idealized model due to fabrication error.

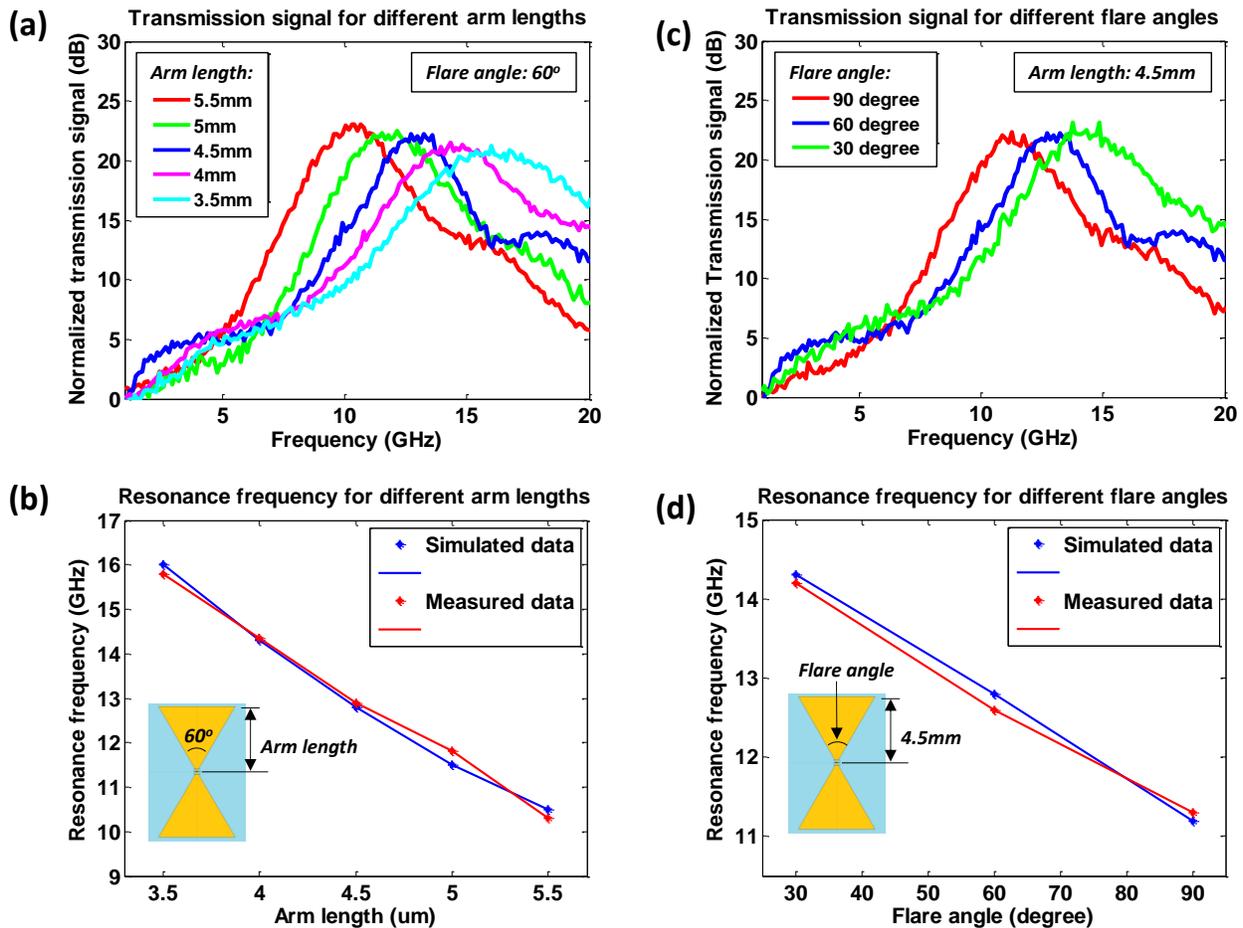

Figure 6. (a) Measured normalized transmission spectrum of bowtie antennas with different arm lengths. (b) Correlation of measured resonant frequency with simulated resonant frequency at different arm lengths. In (a) and (b), flare angles are fixed at 60 degrees. (c) Measured normalized transmission spectrum of bowtie antennas with different flare angles. (d) Correlation of measured resonant frequency with simulated resonant frequency at different flare angles. In (c) and (d), arm lengths are fixed at 4.5mm.

In addition, the far field radiation pattern is measured. The bowtie antenna with arm length of 5.5mm and flare angle of 60 degrees is used in this test. The bowtie antenna is mounted on a rotational stage and is rotated along the x axis in Fig. 1. RF signal from the vector network analyzer is coupled into the bowtie antenna through a GS microprobe. The frequency of this RF signal is set to 10.5GHz which is the resonant frequency of the bowtie antenna. A horn antenna is placed 2m away as a receiving antenna in the far field region. The received power is amplified by an RF amplifier and then measured by a microwave spectrum analyzer. The normalized measured power as a function of rotation angle is shown as a blue curve in Fig. 7. Simulated radiation pattern (red curve) is also overlaid in the figure, showing a good match between simulation and experimental results. This measured radiation pattern indicates dipole-type characteristics of our bowtie antenna. The half power beam width is measured to be about 90 degrees. This wide beam width is good for the antenna to detect electromagnetic waves coming from a large range of incident angles.

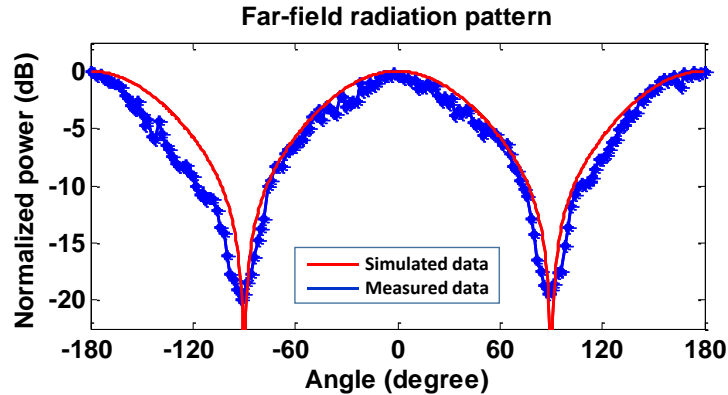

Figure 7. Measured far field radiation pattern (blue) of the bowtie antenna at a frequency of 10.5GHz. Simulated radiation pattern is also overlaid (red).

## 5. CONCLUSION

In summary, we design, fabricate and experimentally demonstrate an integrated broadband bowtie antenna on transparent glass substrate. The bowtie antenna is optimized to cover a broad frequency bandwidth. Numerical simulation shows that, with the arm length of 5.5mm and the flare angle of 60º, the electric field inside the bowtie feed gap can be enhanced by as high as 670 times at 10.5GHz compared to the incident electric field, with a 1-dB RF bandwidth over 9GHz. The dependency of resonant frequency on bowtie geometry, such as arm length and flare angle is numerically computed and experimentally verified. In addition, the radiation pattern of the bowtie antenna is measured, showing a large angular beam width similar to a typical dipole antenna. The bowtie antenna has compact size smaller than 1cm$^2$. This bowtie antenna has potential applications in photonic detection of free-space electromagnetic waves [29-31], RF photonic links and devices [32] complex electromagnetic structures [33], ground penetrating radar [5], THz wave detection [34], plasmonic sensing [35], nano-antenna arrays [8], quantum emitter [36], optical antennas [37], and even light trapping for photovoltaics [38-40]. In addition, from fabrication point of view, solid bowtie antennas [41] or contour bowtie antennas [7] can be fabricated by inkjet printing techniques, which is compatible with roll-to-roll manufacturing processes [42]. These bowtie antennas can also be fabricated on flexible substrates [7, 43] [44] [45]. In addition, the gold material can be replaced by ITO or graphene, together with the transparent feature of glass substrate, to potentially enable some 'invisible' integrated electronic and photonic devices [46]. Furthermore, some bowtie antenna integrated devices previously demonstrated on SOI substrates [22] can be transferred as silicon nanomembranes onto glass substrates [47] or directly fabricated on silicon-on-glass substrates [48] to avoid impacts from backside silicon handles and to enhance their device performance.

## ACKNOWLEDGEMENT

The authors would like to acknowledge the Air Force Research Laboratory (AFRL) for supporting this work under the Small Business Technology Transfer Research (STTR) program (Grant No. FA8650-12-M-5131) monitored by Dr. Robert Nelson and Dr. Charles Lee.